\documentclass[conference,a4paper]{IEEEtran}
\IEEEoverridecommandlockouts
\usepackage[english]{babel}
\usepackage{cite}
\usepackage[hidelinks]{hyperref} % Hyperlinks
\usepackage{amsmath,amssymb,amsfonts}
\usepackage{siunitx}
\usepackage{algorithmic}
\usepackage{graphicx}
\usepackage{textcomp}
\usepackage{xcolor}
\usepackage{siunitx}
\usepackage{booktabs}
\usepackage{comment}
\def\BibTeX{{\rm B\kern-.05em{\sc i\kern-.025em b}\kern-.08em
    T\kern-.1667em\lower.7ex\hbox{E}\kern-.125emX}}
\begin{document}

\title{ Accounting for Subsystem Aging Variability in Battery Energy Storage System Optimization
\thanks{Corresponding author's e-mail: melina.graner@hs-kempten.de}
\thanks{This research is funded by the German Federal Ministry for Economic Affairs and Climate Action (BMWK) via the research project BattLifeBoost (grant number 03EI4068C).}
\thanks{979-8-3315-2503-3/25/\$31.00~\copyright~2025 IEEE.  Personal use of this material is permitted.  Permission from IEEE must be obtained for all other uses, in any current or future media, including reprinting/republishing this material for advertising or promotional purposes, creating new collective works, for resale or redistribution to servers or lists, or reuse of any copyrighted component of this work in other works.}
    }

\author{
    Melina Graner\textsuperscript{1,2},
    Martin Cornejo\textsuperscript{1},    
    Holger Hesse\textsuperscript{2},
    Andreas Jossen\textsuperscript{1} \\[0.1cm]
    \textsuperscript{1}Technical University of Munich,
    TUM School of Engineering and Design, \\
    Department of Energy and Process Engineering,
    Chair of Electrical Energy Storage Technology, Germany\\
    \textsuperscript{2}Kempten University of Applied Sciences; Institute for Energy and Propulsion Technologies, Kempten, Germany \\
}
\maketitle

\begin{abstract}
This paper presents a degradation-cost-aware optimization framework for multi-string battery energy storage systems, emphasizing the impact of inhomogeneous subsystem-level aging in operational decision-making. We evaluate four scenarios for an energy arbitrage scenario, that vary in model precision and treatment of aging costs. Key performance metrics include operational revenue, power schedule mismatch, missed revenues, capacity losses, and revenue generated per unit of capacity loss. Our analysis reveals that ignoring  heterogeneity of subunits may lead to infeasible dispatch plans and reduced revenues. In contrast, combining accurate representation of degraded subsystems and the consideration of aging costs in the objective function improves operational accuracy and economic efficiency of BESS with heterogeneous aged subunits. This fully informed scenario achieves 21\% higher revenue per unit of SOH loss compared to the baseline scenario. These findings highlight that modeling aging heterogeneity is not just a technical refinement but may become a crucial enabler for maximizing both short-term profitability and long-term asset value in particular for long BESS usage scenarios.
\end{abstract}


\section{Introduction}
Battery Energy Storage Systems (BESS) are a key enabler of the energy transition, offering critical flexibility for integrating variable renewable energy sources and supporting grid stability \cite{Hannan.2021}. In Commercial \& Industrial (C\&I) and utility-scale applications, BESS are frequently deployed for energy arbitrage: buying electricity when prices are low and selling or offsetting consumption when prices rise. The efficient operation of such systems is governed by Battery Energy Management Systems (EMS), which utilize optimization-based strategies to generate dispatch schedules \cite{Weitzel.2018}. In arbitrage use cases accurate battery modeling becomes particularly important as revenues depend on precise quantification of dispatchable energy content \cite{Reniers.2021}. A wide spectrum of modeling approaches is available: from abstract bucket models to equivalent circuit models or detailed electrochemical pseudo-two-dimensional (P2D) models \cite{Rosewater.2019}. 
Battery systems are typically composed of multiple cells grouped into subunits or modules, connected in series and parallel configurations. These subunits age at different rates, e.g. due to manufacturing variability, cell balancing issues, and operating conditions, resulting in varying operational behavior of subunits, particularly as the system ages \cite{Barbers.2024}. Both simulations and experimental measurements demonstrate that those aging inhomogeneities among subunits emerge over time, affecting the performance and behavior of the entire system \cite{Paul.2013, Werner.2020}. This phenomenon is even more pronounced in second-life battery systems, where subunits may originate from various sources with diverse usage histories and degradation profiles \cite{Patel.2024}.
Aging-cost-aware optimization gained attention in recent years as a way to increase lifetime and profitability, but most existing models still assume homogeneous aging across the BESS and it's subunits \cite{Collath.2023, Kumtepeli.2024}. This assumption limits the accuracy and effectiveness of the resulting dispatch schedules. This work addresses this gap by emphasizing the need to consider aging inhomogeneities in BESS optimization. We explore how neglecting intra-system variation impacts model accuracy, scheduling quality, profitability, and battery health, thereby making the case for more granular, heterogeneity-aware optimization strategies.

\section{Simulation Framework}
\label{sec:methodlogy}
This work proposes an integrated framework for optimal power scheduling of heterogeneous multi-string BESS, as typically found in the large-scale systems for arbitrage trading.
\subsection{Framework and System Architecture}
\begin{figure}[tb]
    \centering
    \includegraphics[width=1\columnwidth]{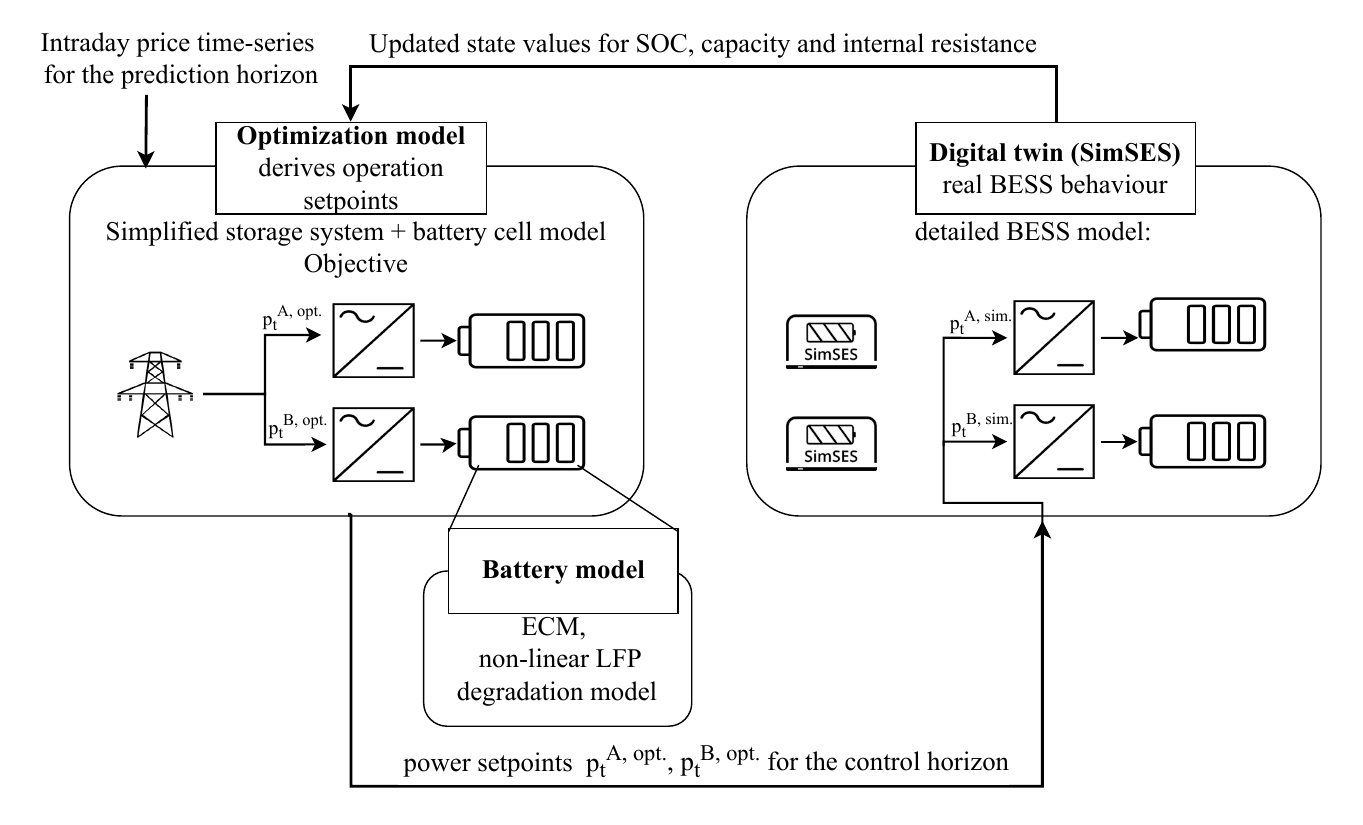} 
    \caption{The proposed simulation framework including the two-string battery model and it's interaction with the digital twin. The two subunits (string A and B) can be optimized and simulated with individual aging characteristics.}
    \label{fig:framework}
\end{figure}

Fig.~\ref{fig:framework} illustrates the proposed simulation framework, consisting of an optimization model and a digital twin of the investigated BESS.
The latter is modeled using the open-source tool SimSES which enables high-fidelity time-series simulations, including equivalent circuit models (ECM), degradation models, and periphery such as AC/DC converters \cite{Moller.2022}. For the cell model, we use a semi-empirical degradation model of a Sony/Murata LFP graphite cell developed in  \cite{Naumann.2018} and\cite{Naumann.2020}. The AC/DC converter efficiency curve is based on \cite{Notton.2010}. Each battery string has a nominal power of 80 kW and an energy capacity of 80 kWh. The optimizer uses simplified models due to computational constraints and solver limitations. The optimization model derives a power dispatch schedule which is passed to the digital twin. The SimSES simulation validates the power setpoints for the control horizon, that lead to updated system states for the state of charge (SOC), remaining capacity and internal resistance. Those parameters alongside the updated price forecasts are the input for the next optimization loop. This process mimics a rolling horizon approach. 
To evaluate economic performance, the framework simulates arbitrage trading on the European intraday market, using real price data from 2021 and assuming perfect foresight.
\subsection{Optimization Model}
\label{optimizationmodel}
The optimization model reflects the batteries internal dynamics and interaction with the grid. The constraints (\ref{eq:ecm-resistance}) - (\ref{eq:ecm-ocv}) characterize the ECM model which defines the relationship between voltage, current, SOC, internal resistance and power, implemented in accordance with the physical behavior of the LFP-battery under investigation \cite{Naumann.2018, Naumann.2020}. Here \( f_{\text{OCV}}(\cdot) \) is a piecewise linear function constructed from \( N \) support points \( \{(\text{SOC}^i, \text{OCV}^i)\}_{i=1}^{N} \). 
The consideration of internal resistance \(R_t\) reflects the quadratic internal battery losses. While the optimization assumes a static average internal resistance, SimSES allows for a temperature and SOC dependent representation of \(R_0\). With aging \(R_0\) increases by the factor \( r^{\text{incr}}\). 
\begin{align}
    R_t &= R_0 \cdot r^{\text{incr}}           & \forall t \in T\label{eq:ecm-resistance} \\
    v_t &= \text{OCV}_t + R_t \cdot i_t          & \forall t \in T \label{eq:ecm-voltage}\\
    -i^{max}  & \leq i_t \leq i^{max}           & \forall t \in T \\
    v^{min}   & \leq v_t \leq v^{max}           & \forall t \in T \\
        p_t^{\text{DC}} &= v_t \cdot i_t        & \forall t \in T \label{eq:ecm-power}\\
    \text{OCV}_t &= f_{\text{OCV}}(\text{SOC}_t) & \forall t \in T \label{eq:ecm-ocv}
\end{align}
The constraints (\ref{eq:soc_eq}) and (\ref{eq:soc_limits}) define the charge-throughput based SOC, where \(Q^{\text{bat}}\) is the updated remaining capacity and \(\Delta t\) the time step duration. It is constrained to the range [0.1, 0.9].
\begin{align}
\text{SOC}_t &= \text{SOC}_{t-1} + \frac{\Delta t}{Q^{\text{bat}}} 
\cdot i_t & \forall t \in T\label{eq:soc_eq}\\
\text{SOC}^{min} & \leq \text{SOC}_t \leq \text{SOC}^{max} & \forall t \in T\label{eq:soc_limits}
\end{align}

The system power \(p_t\) is a combination of charging power \(p_t^{\text{ch}}\) and discharging power \(p_t^{\text{disch}}\), both limited by the system's rated power of 80 kW. A constant inverter efficiency of \(\eta^{inv}\) of 0.95 is assumed for both charging and discharging, following the approach in \cite{Collath.2023}. The digital twin uses a detailed inverter efficiency curve \cite{Moller.2022, Notton.2010}.
\begin{align}
    p_t^{\text{DC}} &= p_t^{\text{ch}} \cdot \eta_{\text{ch}}^{\text{inv}} - (p_t^{\text{disch}} / \eta_{\text{disch}}^{\text{inv}}) & \forall t \in T\\
    p_t &= p_t^{\text{ch}} - p_t^{\text{disch}} & \forall t \in T\\
    0         & \leq p_t^{\text{ch}}, p_t^{\text{disch}} \leq p^{max}   & \forall t \in T                         \label{eq:lp-power-lim}
\end{align}

Based on the LFP cell aging model in \cite{Naumann.2018} and\cite{Naumann.2020}, the constraints (\ref{eq:q_loss_cal}) and (\ref{eq:q_loss_cyc}) capture the effects of calendar and cyclic aging. Here $q_t^{\text{loss, cal}}$ and $q^{\text{loss, cyc}}$ represent incremental relative capacity losses in the given timestep or horizon in per unit. These aging effects contribute to the inherent nonlinearity of the model.
Calendar aging is computed for each timestep t.
\begin{equation}
\label{eq:q_loss_cal}
q_t^{\text{loss, cal}} = 
\frac{
\left( \left( c_1 (\text{SOC}_t - 0.5)^3 + d_1 \right) \! \cdot \! k^{Temp} \right)^2 
}{
2 \cdot \left(1 - \text{SOH}_0^{\text{cal}} \right)}       \! \cdot \!\Delta \text{t}
\end{equation}
Cyclic aging is evaluated over the full prediction horizon.
%Replace \cdot with \! \cdot \! to reduce space before/after the dot:
\begin{equation}
\label{eq:q_loss_cyc}
q^{\text{loss, cyc}} = 
\frac{
\left( (a_2 C^{\text{rate}} + b_2) \! \cdot \! ( c_2 (\text{DOC} - 0.6)^3 + d_2 \right))^2 
}{
2 \cdot(1 - \text{SOH}_0^{\text{cyc}})} \! \cdot \! \Delta \text{FEC}
\end{equation}
Here, \(\text{SOH}_0^{\text{cal}}\) and \(\text{SOH}_0^{\text{cyc}}\) represent the current state of health normalized between 0 and 1. Both are calculated as 1 minus the total accumulated calendar or cyclic losses. They are held constant within a single optimization run and are updated based on the digital twin states between runs. \(\Delta\) FEC, \(C^{\text{rate}}\) and DOC represent the number of full equivalent cycles, the charge/discharge rate, and the depth of cycle. Furthermore \(c_1, d_1, a_2, b_2, c_2\) and \(d_2\) represent fitting parameters for the degradation models described in \cite{Naumann.2018} and\cite{Naumann.2020}. For computational efficiency we choose a simplified charge-throughput based expression of DOC based on \cite{Collath.2023}. By setting DOC in relation to the time passed, we obtain the \(C^{\text{rate}}\) and DOC/2 accounts for the \(\Delta\) FEC.
As state of art C\&I and utility scale battery systems are liquid cooled achieving very homogeneous and near constant temperatures, we confine the simulations to a constant temperature of \(25\,^\circ\text{C}\) which turns the aging stress factor \(k^{Temp}\) into a constant.
\section{Simulation Case Study}
\subsection{Scenarios}
\begin{figure}[bt]
    \centering
    \includegraphics[width=1\columnwidth]{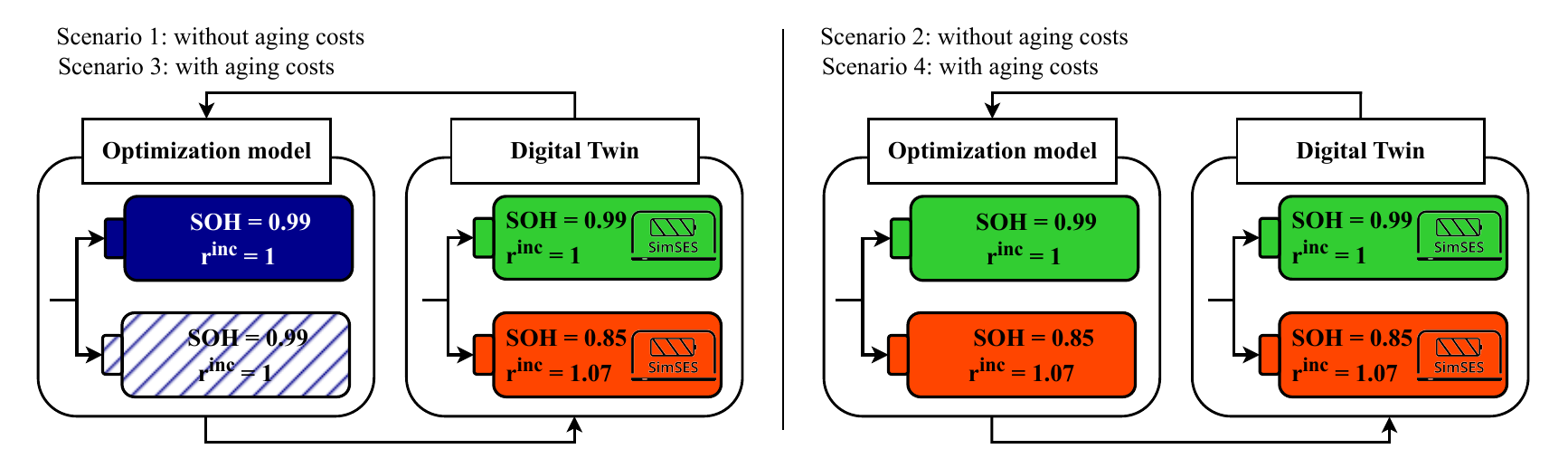} 
    \caption{Depiction of the different model accuracies in the scenarios. String A is considered new, while string B is aged. This heterogeneity of initial SOH and \( r^{\text{incr}}\) is consistently included in the SimSES simulation. Scenarios I and III assume identical conditions for both strings in the optimization, whereas scenarios II and IV incorporate the string-specific degradation states.}
    \label{fig:scenarios}
\end{figure}

The proposed framework (see Section~\ref{sec:methodlogy}) is implemented in Python using Pyomo and the BONMIN solver to simulate four scenarios. All scenarios model a BESS consisting of two parallel battery strings (see Fig.~\ref{fig:scenarios}) operating for one year. A 12-hour prediction horizon, 4-hour control horizon, and 5-minute timestep are used. The start SOC is 50\%.

In \textbf{scenario I (Baseline, state of the art in industrial BESS)}, the EMS does not differentiate between strings. The optimization model uses identical SOH and internal resistance values for both strings resulting in an equal power allocation, while the simulation reflects actual degradation differences.
In \textbf{scenario II (Heterogeneity-Aware)} the optimization model accounts for the aged state of string~B, allowing for more accurate representation and individual power allocation for both strings. Scenarios I and II focus exclusively on generating revenues from the intraday market, where \(c_t^{\text{id}}\) is the intraday market price at each timestep. Their objective function is:
\begin{equation}
\label{eq:LP_obj_1}
\min 
\sum_{t \in T}  p_t \cdot c_t^{\text{id}} \cdot \Delta t.
\end{equation}
To avoid excessive cycling, we set a limit of 2 FEC per day. Without this constraint the optimizer would exploit all revenue opportunities, resulting in an unrealistic high number of FEC (170 per month) which would likely violate typical warranty requirements and induce early capacity fade.

In \textbf{scenario III (Degradation-Cost-Aware Optimization)} the optimization model again assumes identical strings but includes degradation costs in the objective function. In \textbf{scenario IV (Fully Informed)} the optimization model considers both the heterogeneity and degradation costs. For scenarios III and IV, the objective function incorporates battery aging costs \(\mathbb{C}^{\text{ag}}\):
\begin{equation}
\label{eq:LP_obj_2}
\begin{aligned}
\min \quad & \left(\sum_{t \in T}  p_t \cdot c_t^{\text{id}} \cdot \Delta t \right) + \mathbb{C}^{\text{ag}}, \quad \text{with} \\
\mathbb{C}^{\text{ag}} =\; & q^{\text{loss, cyc}} \cdot c^{\text{aging}} \cdot Q^{\text{nom}}/(1 - \text{EOL}).
\end{aligned}
\end{equation}
Here, \(c^{\text{aging}}\) is the cost per lost capacity, \(Q^{\text{nom}}\) the nominal capacity and \(\text{EOL}\) the end of life threshold of 80\% SOH.

\subsection{Metrics}
To assess and compare the defined simulation scenarios, we employ a set of quantitative performance metrics. 
\begin{itemize}
    \item $S^{\text{mismatch}}$: Measures the relative mismatch between scheduled and actual power. Lower values indicate better schedule adherence and model accuracy. 
\begin{equation}
S^{\text{mismatch}} = 1 - \bigl( \sum_{t \in T} p_t^{\text{sim}} \big/ \sum_{t \in T} p_t^{\text{opt}} \bigr)
\end{equation}
    \item $\mathbb{R}$: Higher revenues reflect more effective energy trading and storage use in response to dynamic price signals.
    \begin{equation}
\mathbb{R} = \sum_{t \in T}  p_t \cdot c_t^{\text{id}} \cdot \Delta t 
\end{equation}  
    \item $\Delta$ SOH: The loss of capacity-based SOH over the operation time that indicates aging.
    \item Missed revenues $\mathbb{R}^{\text{missed}}$: Quantifies the relative revenues lost due to discrepancies between the optimized and the actually executed power setpoints. A lower value reflects better alignment between planned and realized operation.
    \begin{equation}
\mathbb{R}^{\text{missed}} = \sum_{t \in T} \left( \mathbb{R}_t^{\text{sim}} - \mathbb{R}_t^{\text{opt}} \right) \big/ \mathbb{R}
\end{equation}    
    \item Revenues per unit of SOH loss $\frac{\mathbb{R}}{\Delta \mathrm{SOH}}$: Higher values indicate more profit per unit of capacity loss and more efficient operation.
\end{itemize}   

\section{Results and Discussion}
\label{section:operational impact}
\begin{figure}[tb]
    \centering
    \includegraphics[width=1\columnwidth]{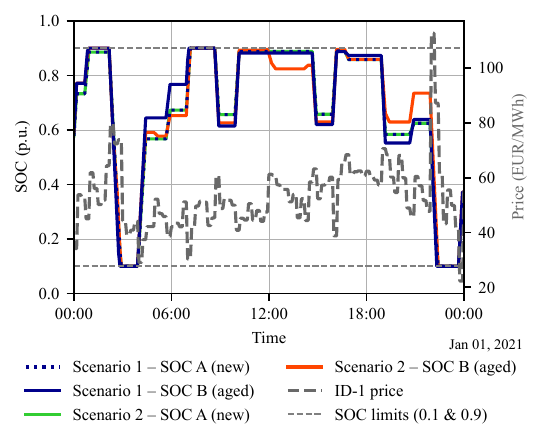} 
    \caption{Exemplary day of operation: SOC trajectories for strings A and B alongside the intraday price signal. While the SOC of string A (dotted blue and green) remains consistent in both scenarios, the SOC of the aged strings B (blue and orange) differs due to different model accuracies.}
    \label{fig:soc and price}
\end{figure}

This section presents and analyzes the results of the four scenarios. We examine how subsystem inhomogeneities affect operational behavior, revenue generation, and degradation in BESS. Furthermore, we evaluate the impact of incorporating aging costs into the optimization objective. Fig.~\ref{fig:soc and price} illustrates typical system behavior: revenues are generated by charging at low and discharging at high prices. 
\subsection{Enhanced Schedule Reliability Through Accurate Aging Representation}
The optimization model in scenario I assumes identical characteristics for both strings, failing to account for the aged condition of string B. Consequently, the model assigns uniform power setpoints to both strings, disregarding critical differences in internal resistance, usable capacity, and SOH.
Therefore string B reaches voltage or SOC limits earlier than expected by the optimization, resulting in a technically infeasible operation plan. String B exhibits significantly higher discrepancies between the scheduled and executed energy throughput, quantified by the power schedule mismatch in Table~\ref{table:results_scenarios_1_and_2}. Consequently, string B experiences frequent underdelivery during discharge events and premature termination of charging cycles. These non-fulfillment events further degrade performance by reducing the actual energy traded during profitable periods, as discussed in the next section.
In contrast, scenario II incorporates the degraded characteristics of string B into the optimization process. The resulting schedule reflects reduced usable capacity and adjusted voltage thresholds. Consequently, the actual power throughput of string B in scenario II remains aligned with the optimization plan, and the schedule mismatch is significantly reduced.
This demonstrates that failing to account for intra-system heterogeneity leads to over-ambitious schedules and reduced fulfillment. 
\begin{table}[htbp]
\caption{Results: scenarios I and II}
\label{table:results_scenarios_1_and_2}
\centering
\renewcommand{\arraystretch}{1.3} % Adjust row height
\resizebox{\columnwidth}{!}{%
\begin{tabular}{|l||c|c||c|c|}
\hline
\textbf{Metric} & \multicolumn{2}{c||}{\textbf{scenario I}} & \multicolumn{2}{c|}{\textbf{scenario II}} \\
\cline{2-5}
 & \textbf{String A} & \textbf{String B} & \textbf{String A} & \textbf{String B} \\
\hline\hline
Power schedule mismatch & 3.7\% & 9.7\% & 3.7\% & 3.6\% \\
\hline
Revenues (EUR) & 4\,846 & 4\,445 & 4\,847 & 4\,941 \\
\hline
Missed revenues & -0.9\% & -1.6\% & -0.9\% & -0.1\% \\
\hline
$\Delta$SOH & 5.1\% & 1.4\% & 5.1\% & 1.3\% \\
\hline
Revenue per unit SOH loss (EUR/$\Delta$SOH) & 93\,630 & 336\,576 & 93\,647 & 375\,475 \\
\hline
\end{tabular}%
}
\end{table}

\subsection{Profitability Gains Through Accurate Aging Representation}
Accurately capturing subsystem-specific aging characteristics in the optimization model directly improves BESS profitability, as scenario II demonstrates. 
While the new string A is accurately modeled in both scenarios and thus generates nearly identical revenues, significant differences emerge in the performance of the aged string B. In scenario I, incomplete fulfillment of scheduled operations reduces the ability of string B to respond to favorable price signals. This mismatch lowers overall revenue potential (see the increased percentage of revenue shortfalls in Table~\ref{table:results_scenarios_1_and_2} for string B).
Scenario II, by contrast, includes a more accurate model of the degraded string B which ensures optimal utilization of the aged string’s available capacity, allowing it to participate more effectively in the energy market. As a result the combined revenues per unit of SOH loss in scenario II are 9\% higher compared to scenario I (469\,122 EUR vs. 430\,203 EUR).
\begin{comment}
Interestingly, string A obtains slightly less revenues than string B in scenario II. We attribute this to higher energy imbalances in discharge direction (6\% for string A vs. 2\% for string B). Because revenue is primarily driven by discharging at peak prices, even small mismatches during discharge periods affect revenues.
\end{comment}
This analysis highlights that by neglecting aging heterogeneity of subunits, optimization schedules may be infeasible for degraded components, resulting in unexploited market opportunities and reduced system efficiency.

\begin{figure}[tb]
    \centering
    \includegraphics[width=1\columnwidth]{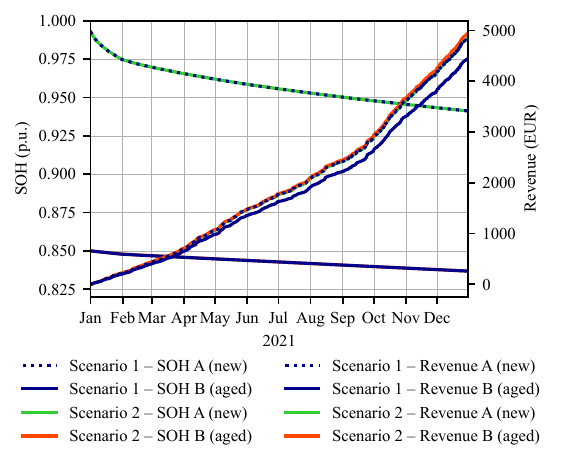} 
    \caption{Cumulative revenues and SOH degradation of strings A and B in scenarios 1 and 2. While string A performs similarly in both cases due to consistent modeling, string B in scenario II benefits significantly from accurate aging representation, achieving higher revenues with slightly lower degradation compared to string B in scenario I.}
    \label{fig:revenues_soh_1-2}
\end{figure}

\subsection{Effect of Model Precision on Battery Degradation}
In scenario I, the model incorrectly treats the aged string B like a new component. This results in higher cycling stress and more frequent operation near voltage and SOC limits, both of which are known to accelerate degradation. 
The total difference in SOH loss over one year between scenarios I and II is modest. As shown in Fig.~\ref{fig:revenues_soh_1-2} and Table~\ref{table:results_scenarios_1_and_2}, string A shows identical aging in both cases and string B experiences only a slightly lower SOH loss in scenario II (1.3\% vs. 1.4\%). Although small, this difference in SOH loss reflects avoidable inefficiencies that could compound over longer lifetimes. Neglecting aging inhomogeneities therefore not only reduces short-term profitability but might also compromise the system's long-term value retention. The following section further addresses this.

\subsection{The Impact of Considering Aging Costs in the Objective Function}
\begin{figure}[tb]
    \centering
    \includegraphics[width=1\columnwidth]{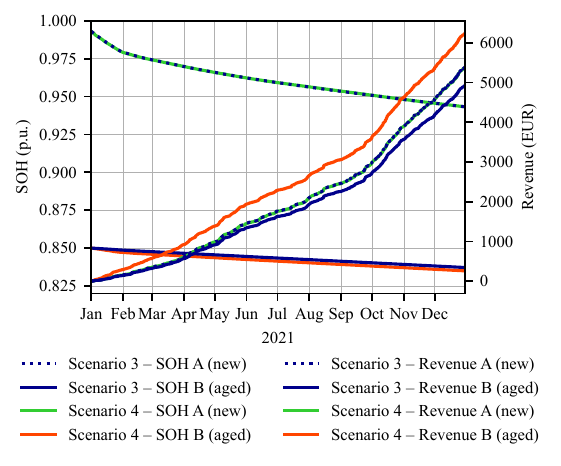} 
    \caption{Cumulative revenues and SOH degradation of strings A and B in scenarios III and IV. Scenario IV shows higher revenues for the aged string B (orange) due to the combined effect of aging-cost-aware optimization and accurate subsystem modeling. Scenario~III, lacking this precision, shows lower performance despite using the same objective function.}
    \label{fig:revenues_soh_3-4}
\end{figure}

In scenarios III and IV, battery degradation is incorporated into the optimization objective. This approach aims to reduce unnecessary wear by selectively following market price signals only when the economic return justifies the associated aging. No fixed limit on FEC per day is imposed, giving the model full flexibility to decide when cycling is economically optimal. The model is supposed to identify a reasonable amount of cycles per day based on the constraints induced by cyclic aging and its impact in the objective function instead of operating between hard limits. To avoid overly conservative operation in the early lifecycle, caused by the square-root dependency of the degradation terms (\ref{eq:q_loss_cal}) and (\ref{eq:q_loss_cyc}), the objective function accounts for cyclic aging losses only. Notably, including the aging costs increases the computational complexity by factor 10.
The results in Table~\ref{table:results_scenarios_3_and_4} and Fig.~\ref{fig:revenues_soh_3-4} show a clear improvement over scenarios I and II. In scenario III, both strings achieve higher revenues while maintaining slightly reduced SOH losses compared to their counterparts in scenario I. This stems from greater operational freedom, due to the absence of hard cycling limits, combined with aging cost penalties that discourage economically unjustified wear. However, the aged string B in scenario III still shows increased schedule mismatch (8.7\%) and increased revenue shortfalls, since the optimizer does not have precise information about its degraded state. In contrast, scenario IV benefits from both aging-cost-aware optimization and accurate representation of subsystem degradation, reducing the power schedule mismatch to 1.8\% and the missed revenues to \text{-0.6\%} for string B. In fact, string B’s performance shows the model's ability to leverage lower marginal aging costs for more frequent and profitable operations, despite slightly higher aging. This is reflected in the highest revenue per unit of SOH loss (415\,012 EUR). The clear advantage of scenario IV over scenario II, which used the same aging state information but excluded aging costs, highlights tangible advantages of integrating both cost and state information into operational decision-making. 

\begin{table}[htbp]
\centering
\caption{Results for scenarios III and IV}
\label{table:results_scenarios_3_and_4}
\resizebox{\columnwidth}{!}{%
\renewcommand{\arraystretch}{1.3} % Adjust row height
\begin{tabular}{|l||c|c||c|c|}
\hline
\textbf{Metric} & \multicolumn{2}{c||}{\textbf{scenario III}} & \multicolumn{2}{c|}{\textbf{scenario IV}} \\
\cline{2-5}
 & \textbf{String A} & \textbf{String B} & \textbf{String A} & \textbf{String B} \\
\hline\hline
Power schedule mismatch & 4.4\% & 8.7\% & 4.4\% & 1.8\%\\
\hline
Revenues (EUR) &  5\,387 & 4\,926 & 5\,379 & 6\,242 \\
\hline
Missed revenues &-3.4\%  & -6.3\% &  -3.4\% & -0.6\% \\
\hline
$\Delta$SOH & 5\% & 1.3\% & 5\% & 1.5\% \\
\hline
Revenue per unit SOH loss (EUR /$\Delta$SOH) & 108\,229 & 384\,100 & 108\,090  & 415\,012 \\
\hline
\end{tabular}%
}
\end{table}
\section{Conclusion and Outlook}
This study demonstrated, through a comparative simulation of four operational scenarios, that neglecting aging inhomogeneities among subunits in BESS leads to technically infeasible schedules, reduced fulfillment, and missed revenue opportunities.
Accurate modeling of degraded subsystems, as conducted in scenario~II, substantially improved both profitability and operational robustness. In fact, scenario~II achieved 9\% higher revenue per unit of SOH loss compared to scenario~I. 
Incorporating aging costs into the optimization objective (scenarios III and IV) further enhanced system performance by enabling more strategic cycling behavior.
Scenario~IV, which combined cost-aware scheduling with detailed subsystem modeling, yielded the best overall results with 21\% higher revenue per unit of SOH loss compared to scenario I (523\,102 EUR vs. 430\,203 EUR). Ultimately, this study highlights that aging-cost-aware and heterogeneity-aware optimization is not only a technical refinement but a driver of long-term economic value in BESS operation.

While the one-year operation results seem promising, the optimizer's tendency to favor cycling the less costly, more aged subsystem may increase SOH imbalance between strings over time. This divergence could lead to an earlier end of life for the entire system. Future research should explore strategies that incorporate SOH balancing objectives or adaptive cost terms to ensure more uniform aging across subsystems. Such refinements may unlock additional lifecycle value in long-duration applications or unbalanced second-life systems.

\bibliographystyle{ieeetr}
\begin{small}  % Use any desired font size here: \small, \footnotesize, \scriptsize
    \bibliography{references_ISGT25}
\end{small}

\end{document}